\documentclass[a4paper,twoside]{article}

\usepackage{amsmath,epsfig}
\usepackage{amssymb,amsfonts}
\usepackage{latexsym}

\newcommand{\Fc}{\mathcal{F}}
\newcommand{\Gc}{\mathcal{G}}

\begin{document}

\title{Modified non-local gravity}

\author{\bf Alexey S. Koshelev\footnote{Postdoctoral researcher of
FWO-Vlaanderen} \\
\it Vrije Universiteit Brussel, Pleinlaan 2, 1050 Brussels, Belgium\\
E-mail: Alexey.Koshelev@vub.ac.be}

\date{}

\maketitle

\begin{abstract}
  In this note I provide an extended version of the talk given at BW2011
workshop. The concise introduction to the non-local SFT motivated models is
given with an emphasis on the non-local generalization of gravity. A number of
open questions and future directions in the development of such models is
outlined.
\end{abstract}

    \section{Introduction}

String-inspired non-local field theory models have gained popularity in recent
years. These models contain an infinite series of higher derivative terms in
their action (and hence the term ``non-local''\footnote{The field equations can
also often be cast as integral equations, a signature of non-local theories.})
similar to what one finds in the action
of the tachyon in string field theory (SFT)~\cite{sft_review}, in stringy toy
models such as p-adic string models~\cite{padic_st} or strings on random
lattice~\cite{random}. There are several motivations for studying such
field theories.
These theories often present novel cosmological features,
and has been used to obtain inflation, dark energy, and bounce. Like
most other higher derivative models, these field theories are better behaved in
the UV as compared to the usual renormalizable models, the quantum theory, in
fact, is finite and provides deviations in scattering cross-sections which may
be detectable in particle physics experiments.
Their study can provide
insight into the ``real'' string theory, for instance the existence of thermal
duality, previously only conjectured using stringy arguments,  was explicitly
demonstrated. Furthermore even though one has an
infinite series of higher derivative terms, these non-local models can avoid the
problem of ghosts by avoiding additional poles in the propagator, and thus
admit the possibility of a consistent UV completion.

Although mostly non-local scalar field theories were explored,
in this note I mainly focus on the non-local modifications of gravity
that was essentially proposed to see whether the big bang problem gets resolved
in these models. While a specific exact nonsingular bouncing solution was indeed
discovered in~\cite{BMS}, in~\cite{BKM} it was further shown that such bouncing
solutions exist for a large class of these non-local models, and that a subset
of these theories admit for a late-time de Sitter expansion (we also note
works \cite{emails} which discuss related models as well). This therefore
provides us with a natural possibility of constructing a past geodesically
complete model of inflation where  the inflationary phase is preceded by a
contracting phase and a nonsingular bounce. In this context we note that it has
been known for a while that the standard near exponentially expanding space-time
is not past geodesically complete.

I will start with a very brief introduction in the non-local SFT motivated
models and then
proceed with more involved non-local set-ups. 

\section{Non-local models basics}

Non-local scalar field action can be easily seen in the computation in SFT.
Consider cubic SFT which may be either bosonic or fermionic.
Performing string world-sheet calculations in the SFT action one arrives to the
action for space-time fields. This action consists of two parts: the quadratic
form which is the kinetic part and the cubic interaction among all the fields.
Schematically we can write it as
\begin{equation}
S=\int
d^Dx\sqrt{-\eta}\left(\frac12\phi_iK_{ij}(\Box)\phi_i-v_{ijk}(e^{-\frac{\beta}
2\Box}\phi_i)(e^{-\frac{\beta}2\Box}\phi_j)(e^{-\frac{\beta}2\Box}\phi_k)\right)
\label{action0}
\end{equation}
where $D$ is the dimension in which our theory lives, $\beta$ is a parameter
determined exclusively by the conformal field theory and $\Box$ is the
d'Alambertian operator computed on the space-time metric. It is natural to
expect $\beta<0$ corresponding to a convergent propagator at large momenta.

Here we see the very important feature: we have explicit non-locality in the
action. We stress here that this is general feature of the SFT based
models\footnote{Appearence of higher derivatives is not unique for this theory.
Non-commutative theories, for instance, also have higher derivative, but these
non-local structures are very different.}. Without the interaction term action
(\ref{action0}) is the free string spectrum and if everything is correct with
SFT then the spectrum coincides with the first quantized string theory. For
cubic SFT this is the case. The form $K_{ij}$ is the first degree polynomial in
the box, meaning it has no higher derivatives. This is a very important
statement saying that our theory has no extra excitations.

If only one field is relevant (like in the tachyon condensation scenarios) one
can quite familiarly analyze properties of a potential. Indeed, defining the
potential as minus action evaluated at $\Box=0$ one can study its vacuum
structure. What is novel and quite non-trivial is the dynamics even for
linearized models.

Action with one scalar field looks like
\begin{equation}
S=\int d^Dx\sqrt{-\eta}\left(\frac12\phi(\Box-m^2)\phi-v(\Box,\phi)\right)
\label{action2}
\end{equation}
where $v$ as before is not neccessarily the cubic monomial in $\phi$ and without
quadratic in $\phi$ term.
Suppose there are two distinct vacua, say $\phi=0$
and $\phi=\phi_0$. Perturbative vacuum has only one excitation with $-k^2=m^2$.
Physics in the non-perturbative vacuum $\phi=\phi_0$ is transparent after the
shift $\phi=\phi_0+\chi$.
The quadratic part becomes
\begin{equation}
S_0=\int d^Dx\sqrt{-\eta}\left(\frac12\chi(\Box-m^2)\chi-\frac{\lambda}2\chi
\Gc(\Box)\chi\right)=\int d^Dx\sqrt{-\eta}\frac12\chi\Fc_\chi(\Box)\chi
\label{action20}
\end{equation}
where we integrated by parts to move the box. $\Gc(\Box)$ is not obviously
expected to be an analytic function and unlikely just an exponent.
With the
general function $\Gc(\Box)$ it is possible that function $\Fc_\chi(\Box)$ has
finite number of poles (giving finite number of ghosts) or just no ghost or
something else.

Indeed, if $\Gc(\Box)=(\epsilon\Box-m_0^2)e^g(\Box)-\Box+m^2$ with $g(\Box)$ an
entire function we end up with $\Fc(\Box)=(\epsilon\Box-m_0^2)e^g(\Box)$. This
produces a single massive excitation which is either ordinary field or ghost
depending on the sign of $\epsilon$.

The new feature is that physics in different vacua can be completely
different in contrast to canonical field theories where at most masses get
shifted. For example, there are may be different number and nature of states
in distinct vacua. The more comprehensive analysis can be found in, for
instance,~\cite{KAV} and references therein.

\section {Non-local gravity}

In fact nothing restricts us to scalar fields only and one can build models
with other fields involving towers of d'Alambertian operators. We therefore
proceed to the non-local gravity.

The non-local modification of the Einstein gravity, which is proposed in
\cite{BMS} is described by the following action:
\begin{equation}
 S=\int d^4x\sqrt{-g}\left(\frac {M_P^2}{2}R+\frac{1}{2}R\Fc(\Box/M_*^2)R-\Lambda\right)
 \label{nlg_action}
\end{equation}
where   $M_P$
is the Planck mass: $M_P^2=1/(8\pi G_N)$, $G_N$ is the Newtonian gravitational constant,
 $M_{\ast}$ is the mass scale at which the higher derivative
terms in the action become important.
An analytic function $\Fc(\Box/M_*^2)=\sum\limits_{n\geqslant0}f_n\Box^n$ is an
ingredient inspired by the SFT. The operator $\Box$ is the covariant d'Alembertian.
In the case of an infinite series we have a non-local action.

Such a model exhibits a number of interesting properties outlined in the
introduction for the non-local modifications of gravity. Namely, one can
construct solutions to the equations of motion which describe a non-singular
bounce and/or de Sitter late time attractor. The surprising point is that exact
analytic solutions can be found.

On the other hand an intriguing connection with $p$-adic like models arises if
one introduces a scalar field in order to rewrite the action. Indeed, one can
check that the following equivalent action can be written:
\begin{equation}
 S=\int d^4x\sqrt{-g}\left(\frac
{M_P^2}{2}R(1+\frac
2{M_P^2}\psi)-\frac{1}{2}\psi\Gc(\Box/M_*^2)\psi-\Lambda\right)
 \label{nlg_action_psi}
\end{equation}
Varying over $\psi$ and substituting the resulting solution of the
corresponding equation of motion back in the action restores
(\ref{nlg_action}) provided $\Gc=\Fc^{-1}$.

Further one can make the conformal transformation and rescale the metric such
that there is no non-minimal coupling. This is however not as smooth as in
local theories because the box is the covariant box and the rescaling of the
metric will introduce non-trivial dependence on the scalar field inside the
function $\Fc$. It still would be feasible to analyze the linearized theories
around a given vacuum but the whole model would look much more cumbersome.

The above quadratic in scalar curvature actions without the cosmological
constant are almost enough to study the Minkowski background\footnote{See
\cite{p2} for the most general non-local action suitable for the Minkowski
space-time.}. Indeed, higher curvature terms would vanish both in the background
and in the linear variation of equations of motion producing no new effects.

One can check, however, that in order to have the de Sitter solution one must
keep the cosmological constant explicitly. It is easy to understand since de
Sitter metric produces covariantly constant Riemann tensor and therefore higher
derivatives are not involved while it is known that $R^2$ theories of gravity
do not have the de Sitter solution. To overcome this higher degrees of curvature
may be introduced in the action. This can make it possible to have the de
Sitter Universe without having the cosmological term explicitly.

The major and significant difference with well known modified theories of
gravity is the fact that higher derivative structures can eliminate unwanted
(ghost) excitations keeping the theory well defined.

\section{Simplest proposal to avoid the cosmological constant}

As one of the simplest proposal we promote action (\ref{nlg_action}) to the
following form
\begin{equation}
 S=\int d^4x\sqrt{-g}f(\Pi(\Box/M_\star^2)R)
 \label{nlg_action_f}
\end{equation}
It is clearly just action (\ref{nlg_action}) if $f$ is the quadratic
polynomial, $\Pi^2=\Fc$ and constants are adjusted properly.
Again, almost as in usual $f(R)$ gravity theories one can pass to the kind of
scalar-tensor theory which is now non-local though.

Canonically in local modified gravity theories
\begin{equation*}
 S=\int d^4x\sqrt{-g}f(R)\to S=\int d^4x\sqrt{-g}(Rf'(\psi)-\psi
f'(\psi)+f(\psi))
\end{equation*}
where prime denotes the derivative w.r.t. the argument.
In case of non-local theories
we just would have
\begin{equation*}
 S=\int d^4x\sqrt{-g}((\Pi R)f'(\psi)-\psi
f'(\psi)+f(\psi))
\end{equation*}
Further one can move the operator $\Pi$ from $R$ to $f'(\psi)$ and make the
field redefinition $\Pi f'(\psi)=\chi$. At this stage the action is again
similar to $p$-adic string theory. One more transformation, which is the metric
rescaling is possible to separate explicitly the scalar curvature. In this model
de Sitter solution exists provided there is a non-trivial vacuum for the field
$\chi$.

The next question whether such an action may have again the non-singular
bounce as well, is there a solution which starts with the bounce and ends up in
the de Sitter phase and then one may ask whether perturbations behave good or
not.

This is surely a highly non-trivial thing to construct at least one solution.
The only hope so far is to make use of some ansatz. In case of the quadratic
action (\ref{nlg_action}) the following ansatz does work
\begin{equation}
\Box R=r_1R+r_2
\label{ansatz}
\end{equation}
One can check that specializing to the space-homogeneous and isotropic
spatially flat FRW 4-dimensional Universe with the metric
$ds^2=-dt^2+a(t)^2d\vec{x}^2$ the following scale factors satisfy
the above ansatz
\begin{equation*}
a(t)=a_0\cosh(\lambda t)\text{ and }a(t)=a_0e^{\frac{\lambda}2t^2}
\end{equation*}
In order for this functions become solutions one has to adjust coefficients of
the Taylor expansion of operator function $\Fc$ (or $\Pi$). For the quadratic
action only two conditions on values of $\Fc(r_1)$ and $\Fc'(r_1)$ arise.
At present stage it is an open question to propagate this idea to action
(\ref{nlg_action_psi}) since equations become more complicated.

\section{Beyond scalar curvature}

As another direction one introduce in the action terms containing Ricci and
Riemann tensors like
\begin{equation*}
R_{\mu\nu}\Fc R^{\mu\nu}\text{ and
}R_{\mu\nu\alpha\beta}\Fc R^{\mu\nu\alpha\beta}
\end{equation*}
In a local action in 4 dimensions Riemann tensor squared could be dropped due
to the fact that Gauss-Bonnet term is a topological term. In our case, however,
even in 4 dimensions both such terms make sense because the presence of an
operator in the middle renders them non-trivial. Action exactly of this type is
considered in \cite{p2} and looks like
\begin{equation*}
S=\int d^4x\sqrt{-g}\left(\frac{M_P^2}2R+a_1R\Fc_1 R+a_2R_{\mu\nu}\Fc_2
R^{\mu\nu}+a_3R_{\mu\nu\alpha\beta}\Fc_3 R^{\mu\nu\alpha\beta}+\Lambda\right)
\end{equation*}
such an action may be a very interesting generalization, it also surely admits
inclusion of higher degrees of curvature tensors but the analysis is highly
non-trivial because of the fact that covariant derivatives acting on second and
fourth rank tensors involve the corresponding number of metric connections thus
making the variation w.r.t. the metric very tedious.

The primary question which must be answered here is the formulation of the
ghost free condition in a non-trivial (i.e. non-Minkowski) background.
Also it may be interesting to find whether it is possible to reformulate the
latter action in a simpler way by means of scalar fields like it was done in
the previous Section in case of only the scalar curvature.

\section{Outlook}

Even though we understand how the background solutions can be constructed in
such non-local models and really can tame them in various regimes and even
construct exact analytic solutions in specific cases there are major issues one
must analyze one by one in each particular setup.

First it is important to guarantee that the perturbative spectrum does not
contain ghosts. This is exactly the place where non-local operators do their
job. Second, one must answer the question are the perturbations well
behaved? To do this the best way is to compute the second variation of the
action but this task is quite difficult for non-trivial (i.e. non-constant
curvature) solutions. So far it seems possible to do this either for Minkowski
or de Sitter backgrounds. Third and not yet explored is the question: do the
obtained solutions represent general behavior of a model or they are just
specific measure zero configurations.

There is a  natural hope that reformulation of non-local gravity models in
terms of $p$-adic like action may shed more light since $p$-adic theories are
more explored to the moment. It would be interesting to see more specifically
the correspondence in between of these two classes of models.  On this way
action (\ref{nlg_action_f}) looks as the most doable forthcoming project in
developing non-local models since it has clear reformulation in terms of an
additional  scalar field. A detailed analysis of models based on action
(\ref{nlg_action_f}) is the goal of the upcoming paper \cite{progress}.

\section*{Acknowledgments}
I would like to thank organizers of the BW2011 workshop for
inviting me to this  meeting, for creating the very stimulating
environment during the time of the conference and the very warm hospitality.
This work is supported in part by
the Belgian Federal Science Policy Office through the Interuniversity
Attraction
Pole P6/11, by the “FWO-Vlaanderen” through the project G.0114.10N,
by the RFBR grant 11-01-00894-a, and by the 
grant of the Russian Ministry of Education and Science NSh-4612.2012.1.

    \end{document}